\documentclass[a4paper,11pt]{article}
\usepackage{pos}
\usepackage{graphicx}
\usepackage{natbib}
\usepackage{amssymb}
\usepackage{amsmath}
\usepackage{xspace}
\usepackage{multicol}
\usepackage[normalem]{ulem}
\usepackage{color,soul}

\graphicspath{{./}{figs/}}

\title{Non-thermal emission from young supernova remnants in dense circumstellar environments}
 \ShortTitle{Gamma-ray emission from young SNRs remnants in dense CSM}

\author[a,b]{Robert Brose}
\author*[a,c]{Jonathan Mackey}
\author[d,e]{I. Sushch}

\affiliation[a]{Dublin Institute for Advanced Studies,
  31 Fitzwilliam Place, Dublin 2, Ireland}

\affiliation[b]{Institute of Physics and Astronomy, University of Potsdam,
14476 Potsdam, Germany}

\affiliation[c]{Centre for AstroParticle Physics and Astrophysics (CAPPA), DIAS Dunsink Observatory\\
Dunsink Lane, Dublin 15, Ireland}

\affiliation[d]{Centre for Space Research, North-West University, 2520 Potchefstroom, South Africa}

\affiliation[e]{Astronomical Observatory of Ivan Franko National University of Lviv, Kyryla i Methodia 8, \\79005 Lviv, Ukraine}

\emailAdd{broserob@cp.dias.ie}
\emailAdd{jmackey@cp.dias.ie}

\abstract{
Supernova remnants are known to accelerate cosmic rays (CRs) on account of their non-thermal emission of radio waves, X-rays, and gamma rays. However, the ability to accelerate CRs up to PeV-energies has yet to be demonstrated. The presence of cut-offs in the gamma-ray spectra of several young SNRs led to the idea that PeV energies might only be achieved during the very initial stages of a remnant’s evolution.
We use the time-dependent acceleration code RATPaC to study the acceleration of cosmic rays in supernovae expanding into dense environments around massive stars, where the plentiful target material might offer a path to the detection of gamma-rays by current and future experiments.
We performed spherically symmetric 1-D simulations in which we simultaneously solve the transport equations for cosmic rays, magnetic turbulence, and the hydrodynamical flow of the thermal plasma in the test-particle limit.
We investigated typical parameters of the circumstellar medium (CSM) in the freely expanding winds around red supergiant (RSG) and luminous blue variable (LBV) stars.
The maximum achievable energy might be limited to sub-PeV energies despite strong magnetic fields close to the progenitor star that enhance turbulence-damping by cascading: we find a maximum CR energy of 100-200\,TeV, reached within one month after explosion.
The peak luminosity for a LBV progenitor is $10^{43}$\,erg\,s$^{-1}$ ($10^{42}$\,erg\,s$^{-1}$) at GeV (TeV) energies and, for a RSG progenitor, $10^{41}$\,erg\,s$^{-1}$ ($10^{40}$\,erg\,s$^{-1}$). 
All calculated SNe reach their peak gamma-ray luminosity after $\lesssim1$\,month and then fade at a rate $\sim t^{-1}$, as long as the SN shock remains in the freely expanding wind of the progenitor.
Potentially detectable gamma-ray signals can be expected in the \textit{Fermi-LAT} waveband weeks to months after an explosion into a freely expanding wind. }

\FullConference{37$^{\rm{th}}$ International Cosmic Ray Conference (ICRC 2021)\\
		July 12th -- 23rd, 2021\\
		Online -- Berlin, Germany}

\begin{document}
\maketitle

\section{Introduction}
Even though supernova remnants (SNRs) are considered to be the best candidate sources of the Galactic Cosmic Rays (CRs), evidence that they are able to accelerate CRs up to the required $3\,$PeV is elusive. Observational constraints put the maximum achievable energy at $\approx100\,$TeV at most, whereas young SNRs as Tycho and Casiopeia A show even lower cutoff energies despite their youth \citep{2020ApJ...894...51A}.
Recent studies of CR acceleration suggest that the escape of CRs once SNRs enter the Sedov stage might be responsible for the observed, soft late-time spectra of evolved SNRs \citep{2020A&A...634A..59B, 2019MNRAS.490.4317C}.
The aforementioned remnants are close to entering the Sedov-phase, which makes an acceleration of CRs beyond PeV-energies unlikely in evolved SNRs since the rapidly decreasing maximum energy is the main driver of the particle escape after the onset of the Sedov phase.

It has been proposed instead that very young SNRs with ages below $20\,$years are the only objects capable of achieving PeV energies \citep{2018MNRAS.479.4470M, 2020MNRAS.494.2760C}.
\citet{2013MNRAS.431..415B} calculate that the CR density gradient is only large enough during the initial phases to drive turbulence to relevant scales by the non-resonant streaming instability.
Recently, there has been evidence for $\gamma$-ray emission from very young SNRs at \textit{Fermi-LAT} energies \citep{2020ApJ...896L..33X, 2018ApJ...854L..18Y} whereas a detection at the highest energies remains elusive \citep{2019A&A...626A..57H}. Theoretically, remnants expanding in dense circumstellar media are regarded as the best candidates for acceleration to PeV energies and for a detection of the $\gamma$-ray emission produced by $p$-$p$ collisions.
In this work, we use numerical simulations to make predictions for $\gamma$-ray emission from Type-IIn and Type-IIP SNe associated with luminous blue variable (LBV) and red supergiant (RSG) progenitor stars.

\section{Basic equations and assumptions}
We use the \textbf{R}adiation \textbf{A}cceleration \textbf{T}ransport \textbf{Pa}rallel \textbf{C}ode (RATPaC) to calculate the particle acceleration and subsequent thermal and non-thermal emission. A detailed description of the code can be found here: \cite{2012APh....35..300T, 2016A&A...593A..20B, 2020A&A...634A..59B}. 

\subsection{Cosmic rays}
We model the acceleration of cosmic rays using a kinetic approach in the test-particle approximation. The time-dependent transport equation for the differential number density of cosmic rays $N$ \citep{Skilling.1975a} is given by:

\begin{align}
\frac{\partial N(r,p,t)}{\partial t} =& \nabla(D_r\nabla N(r,p,t)-\mathbf{u} N(r,p,t))
 -\frac{\partial}{\partial p}\left( (N(r,p,t)\dot{p})-\frac{\nabla \cdot \mathbf{ u}}{3}N(r,p,t)p\right)+Q
\label{CRTE}\text{ , }
\end{align}
where $D_r$ denotes the spatial diffusion coefficient, $\textbf{u}$ the advective velocity, $\dot{p}$ energy losses and $Q$ the source of thermal particles.
We inject electrons and protons according to the thermal leakage injection model \citep{Blasi.2005a,1998PhRvE..58.4911M}. 

\subsection{Magnetic field}\label{sec:MF}
We consider two configurations for the large-scale magnetic field in this study.
The first assumes a constant $5\,\mu$G throughout the upstream region and a shock-compressed, uniform $16\,\mu$G field in the downstream.
In the second configuration, we follow the induction equation for the transport of the large-scale field \citep{Telezhinsky.2013}. As an initial condition, we assume a magnetic field originating from the progenitor star, following the prescription by Voelk \& Forman \citep{1982ApJ...253..188V}. For simplicity, we neglect the region where the magnetic-field follows $1/r^2$-dependence close to the progenitor star (which covers a region smaller than $10^{15}\,$cm even for a RSG rotating at $1\,$km\,s$^{-1}$) and thus use
\begin{align}
    B(r) &= B_\ast \frac{r_\ast}{r} \text{ , }
\end{align}
where $B_\ast$ is the field at the surface and $r_\ast$ the radius of the progenitor star.
We use $r_\ast=1000R_\odot$ and $B_\ast=1\,$G in this work.

In addition to the large-scale magnetic field, we consider the contribution from the amplified magnetic turbulence. The total magnetic field is thus given by $B_\text{tot} = \sqrt{B_0^2+B_\text{Turb}^2} \text{ , }$ where $B_\text{Turb}$ is derived by integrating our turbulence spectrum in wavenumber space.

\subsection{Magnetic turbulence}
In parallel to the transport equation for CRs, we solve a transport equation for the magnetic-turbulence spectrum. The temporal and spatial evolution of the spectral energy-density per unit logarithmic bandwidth, $E_w$, is described by
\begin{equation}
 \frac{\partial E_w}{\partial t} +  \cdot \nabla (\mathbf{u} E_w) + k\frac{\partial}{\partial k}\left( {k^2} D_k \frac{\partial}{\partial k} \frac{E_w}{k^3}\right) 
=2(\Gamma_g-\Gamma_d)E_w \text{ . }
\label{eq:Turb_1}
\end{equation}
Here, $\mathbf{u}$ denotes the advection velocity, $k$ the wavenumber, $D_k$ the diffusion coefficient in wavenumber space, and $\Gamma_g$ and $\Gamma_d$ the growth and damping terms, respectively \citep{2016A&A...593A..20B}.

We use a growth-rate based on the resonant streaming instability \citep{Skilling.1975a, 1978MNRAS.182..147B},
\begin{align}
    \Gamma_g &= \frac{v_\text{A}p^2v}{3E_\text{w}}\left|\frac{\partial N}{\partial r}\right| \text{ , }\label{eq:growth}
\end{align}
where $v_\text{A}$ is the Alfv'en-velocity. Classically, Alfv'enic turbulence can amplified to a level, where $\delta B/B_0 \approx 1$. However, amplification beyond $\delta B/B_0 \approx 1$ is assumed to be possible by e.g. the non-resonant streaming instability \citep{2000MNRAS.314...65L, 2004MNRAS.353..550B}. The gradient of the CR-distribution, which also determines the growth of the non-resonant mode, is strong enough to amplify $\delta B$ to values greater than $B_0$ for the young objects considered here.

The growth of the magnetic turbulence and hence the magnetic field is balanced by cascading. This process is described as a diffusion process in wavenumber space, and the diffusion-coefficient is given by \citep{1990JGR....9514881Z, Schlickeiser.2002a}
\begin{align}
    D_\text{k} &= k^3v_\text{A}\sqrt{\frac{E_\text{w}}{2B_0^2}} \text{ . }
\end{align}
This phenomenological treatment will result in a Kolmogorov-like spectrum, if cascading is dominant.

\subsection{Hydrodynamics}
The evolution of a SNR without CR feedback can be described with the standard gas-dynamical equations. This system of equations is solved under the assumption of spherical symmetry in 1-D using the PLUTO code \citep{2007ApJS..170..228M}. It has to be noted, that radiative losses will play an important role in the early evolution of the remnant, especially in a very dense circumstellar medium (CSM). However, photons can not easily escape and a different equation of state has to be used to accurately describe regions in the far-downstream of the forward shock \citep{2019A&A...622A..73O}.

The results of the hydro simulation for the density, velocity, pressure, and temperature distributions are then mapped onto the spatial coordinate of the CR and turbulence grid, respectively. 
The remnants are modeled as CC events using the initial conditions of \cite{1982ApJ...258..790C}. We used an ejecta mass of $M_\text{ej, LBV}=10\, M_\odot$ and $M_\text{ej, RSG}=3\, M_\odot$, with a power-law index of $n_\text{LBV}=10$ and $n_\text{RSG}=9$ for the LBV and RSG cases, respectively. The parameters assumed for the ambient medium are shown in table \ref{tab:performance}.
To improve the computational performance a dynamical re-gridding scheme was introduced, that dynamically moves the grid boundaries together with the shock structures.

\begin{table}[h]
  \centering
  \begin{minipage}{0.59\textwidth}
  \begin{tabular}{|c|c|c|c|}
    \hline
    Model & $\dot{M} [M_\odot/$yr] & $V_\text{sh}$ [km/s]  & $R_\text{sh}$ [cm] \\
    \hline
    \hline
    LBV  & $10^{-2}$ & 100 & $1.2\cdot10^{14}$ \\
    RSG  & $8.3\times10^{-5}$ & 15 & $6\cdot10^{13}$ \\ 
    \hline
  \end{tabular}
\end{minipage}
\begin{minipage}{0.39\textwidth}
\caption{Parameters for the progenitor stars winds and initial remnant sizes} 
  \label{tab:performance}
\end{minipage}
\end{table}

\section{Results}
\begin{figure}
    \centering
    \includegraphics[width=0.95\textwidth]{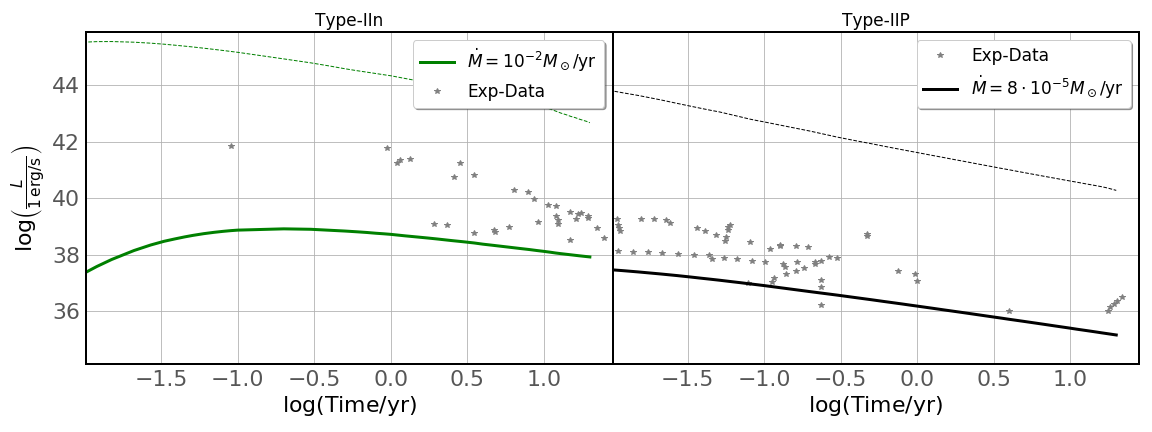}
    \caption{Comparison of the thermal X-ray emission, where the unabsorbed flux is shown in dashed lines and the flux including absorption in solid lines. The experimental data is taken from \cite{2014MNRAS.440.1917D}.}
    \label{fig:ThermalX}
\end{figure}

We followed the evolution of the remnants for 20 years. In order to verify the hydrodynamic evolution of our remnants, we calculated the thermal X-ray continuum emission based on the approximations from Hnatyk \& Petruk \citep{1999A&A...344..295H}. Figure \ref{fig:ThermalX} shows the evolution of the X-ray emission compared to measurements from Type-IIn and Type-IIP SNs respectively.  
We accounted for potential absorption in the remnant itself and in the dense ISM, calculating the optical depth based on the hydrogen-density from our simulation and the cross-section given by \cite{2019MNRAS.486.1094M,1983ApJ...270..119M}.
We are likely slightly overestimating the absorption since we can not considering the ionization-state of the medium. However, the derived X-ray emission is well within the expected range and shows the observed time-evolution.

\subsection{Maximum particle energies}
\begin{figure}
    \centering
    \begin{minipage}{0.59\textwidth}
    \includegraphics[width=0.95\textwidth]{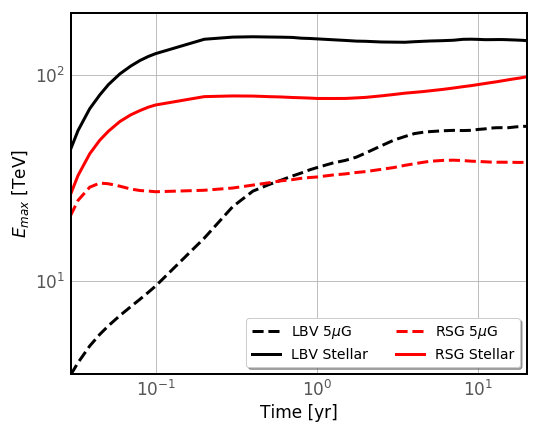}
    \end{minipage}
    \begin{minipage}{0.39\textwidth}
    \caption{Maximum energy of protons for a LBV progenitor (black) and a RSG progenitor (red).
    Solid lines indicate a $1/r$ magnetic field derived from the progenitors surface field and dashed lines indicate models with a constant $5\,\mu$G upstream field.}
    \label{fig:MaxE}
    \end{minipage}
\end{figure}

We derived the time evolution of the maximum energy for our four different configurations. The results are shown in Figure \ref{fig:MaxE}.
It is evident that a higher ambient density is increasing the maximum energy (LBV vs. RSG) since a denser CSM means that more particles get injected into the acceleration process as CRs. A higher CR density translates into more driving of turbulence by the CR gradient in equation (\ref{eq:growth}). In this early stage of the evolution, the shock speed is only weakly depending on the ambient density. The density dependence is stronger in the Sedov stage where the effect of potentially more injected CRs gets almost completely canceled by a correspondingly lower shock speed. 

At the same time, the higher ambient magnetic field is translating into a higher particle energy. In general, the higher field means there is a higher growth of turbulence due to the higher Alfv\'en speed. However, once the level of turbulence exceeds $\delta B/B_0=1$, there is a significant effect from the magnetic turbulence on the background field and the scaling of the cascading term for magnetic turbulence changes 
\begin{align}
    D_k \propto 
    \begin{cases}
        \sqrt{E_\text{w}} &\text{for $E_\text{w} < B_0^2/8\pi$}\\
        E_\text{w}^{3/2} &\text{for $E_\text{w} > B_0^2/8\pi$} \text{ . }\\ 
    \end{cases}
\end{align}
In the strong turbulence regime the cascading rate is increasing faster ($\propto B_\text{tot}^{3/2}$) than the growth-rate ($\propto B_\text{tot}$).
This limits the maximum field to values below that predicted by works just considering the saturation level of the non-resonant instability as the limiting factor for the amplification \citep{2018MNRAS.479.4470M}.
The different acceleration time associated with the different background fields then translates into the observed difference of the maximum energies for the different field configurations. 
The maximum achievable energies are well below the PeV-energies needed for the sources of Galactic CRs but can reach more than $100\,$TeV in the LBV case when a strong ambient stellar field is present.

\subsection{Gamma-ray emission}
\begin{figure}
    \centering
    \includegraphics[width=0.95\textwidth]{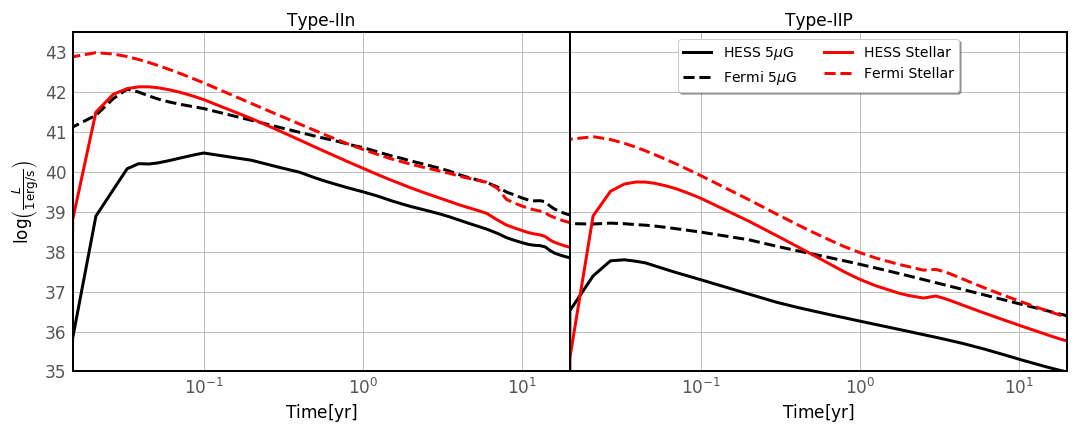}
    \caption{Gamma-ray luminosities in the \textit{Fermi-LAT} energy range (dashed) and \textit{H.E.S.S.} energy range (solid) for a constant ambient field (black) and {stellar} ambient field.}
    \label{fig:GammaL}
\end{figure}

We derived the $\gamma$-ray luminosity in the 1-10 TeV (hereafter \textit{H.E.S.S.}) and 0.1-300GeV (hereafter \textit{Fermi-LAT}) energy bands. Figure \ref{fig:GammaL} illustrates the evolution of the $\gamma$-ray luminosity for our models.
We start the acceleration of CRs 3 days after the explosion. Despite the high growth-rate of turbulence in the stellar-field case, it needs about 5-10 days for the $\gamma$-ray luminosity in the \textit{H.E.S.S.} band to reach its peak value. The peak is reached faster in the LAT band on account of the lower energy of particles radiating at these energies. 
A stronger ambient field translates into a faster initial growth and thus an earlier peak of the $\gamma$-ray luminosity. Likewise, a higher ambient density translates into a higher $\gamma$-ray luminosity. 
In all models, the luminosity is decreasing roughly as $1/t$ after its peak.

The $\gamma$-ray luminosities shown here are neglecting $\gamma$-$\gamma$ interactions and the subsequently dimming of the $\gamma$-ray flux as pointed out by Cristofari et al. \citep{2020MNRAS.494.2760C}.
They showed that the flux for low energy $\gamma$-rays can be efficiently attenuated for at least 10 days after explosion, complicating a detection.
The prediction of the $\gamma$-ray luminosity in the \textit{H.E.S.S.}\ band is in line with current observational upper limits of $\approx10^{40}\,$erg/s \citep{2019A&A...626A..57H}.
We here considered only SNe with the highest mass-loss rates prior to explosion and predict at most $10^{42}\,$erg/s and $10^{40}\,$erg/s for a LBV and RSG progenitor respectively.

\begin{figure}
    \centering
    \includegraphics[width=0.99\textwidth]{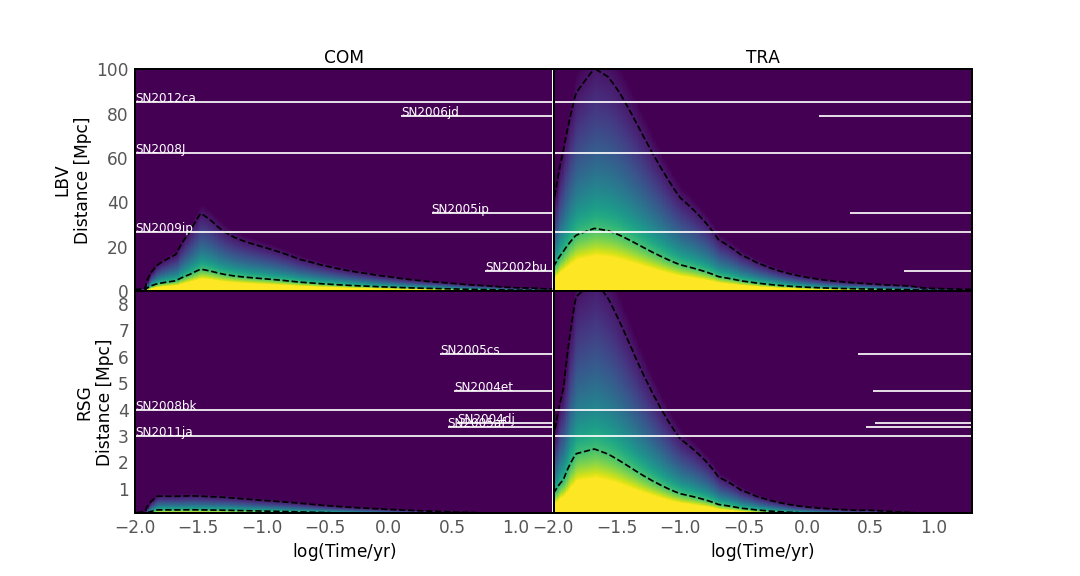}
    \caption{Detectability of nearby SNe in the \textit{Fermi-LAT} band.
        The logarithmic colour scale shows flux as a function of time post-explosion and of distance to the SN, where lighter colours are brighter.
        The lower (upper) dashed line shows \textit{Fermi-LAT} $\approx8\sigma$ detection threshold for 1 day (1 month) of observation.}
    \label{fig:Fermi}
\end{figure}

We started to investigate the prospects for a detection of a $\gamma$-ray signal by \textit{Fermi-LAT}. Unlike the IACTs, \textit{Fermi-LAT} is constantly monitoring the sky and thus allows a retrospective study of potentially interesting objects.
Figure \ref{fig:Fermi} shows how the gammay-ray flux at LAT-energies depends on the distance of the SNR.
The dashed contours illustrate the detection threshold of by \textit{Fermi-LAT} for one day and one month of observation.
The white lines indicate the overlap of the LAT data-set with known SNe of Type-IIn and Type-IIP.
Our prediction is that all supernovae expanding into wind density-profiles are already fading significantly after less than one month, and so the maximum distance out to which a SN can actually be detected is not the extremum of each curve, but rather the $y$-value corresponding to a post-explosion time of 1 day or 1 month.
It can be seen that there is basically no Type-IIP SN that was close enough to be detected by \textit{Fermi-LAT} regardless of the magnetic-field model. There is some overlap for SN2011ja and SN2008bk, however, the emission faded within less than $30\,$days where a full month of elevated flux would have been needed for a detection. Furthermore, our applied mass-loss rate, and thus the $\gamma$-ray luminosity, is at the high end of the expected parameter range.

The situation looks more promising for the Type-IIn SN2009ip, that could be detected within less than a month by \textit{Fermi-LAT} if the acceleration is as efficient as predicted by our stellar-field model. There remains the caveat, that our applied LBV mass-loss rate is about 10 times higher than the value of $10^{-3}M_\odot$\,yr$^{-1}$ that is derived from X-ray observations for SN2009ip \citep{2013ApJ...768...47O}. So far, there is no good candidate for a nearby Type-IIn or Type-IIP SN that should have been detected by \textit{Fermi-LAT}. The situation is complicated by the fact that many SNe occur in star-forming or star-burst galaxies that might already have significant background $\gamma$-ray emission. Thus, a variation of flux imposed by the SN is potentially harder to detect than an isolated, background-free, point source.
Further, as mentioned in context of the $\gamma$-ray luminosities, $\gamma$-$\gamma$ interactions reduce the predicted $\gamma$-ray flux and this still needs to be taken into account in our calculations.

\section{Conclusions}
We performed numerical simulations of particle acceleration in very young SNRs expending in dense circumstellar media, solving time-dependent transport equations of CRs and magnetic turbulence in the test-particle limit alongside the standard gas-dynamical equations for CC-SNRs. We derived the CR diffusion coefficient from the spectrum of magnetic turbulence, that evolves through driving by the resonant streaming instability as well as cascading and wave damping.
\begin{itemize}
    \item We found good consistency between our hydro-models and the observed evolution of the thermal X-ray emission from SNe.
    \item We showed that even for very dense CSM and the presence of strong ambient magnetic fields, the maximum CR energy hardly exceeds $\approx100\,$TeV, limited by efficient damping of magnetic turbulence.
    \item The peak-luminosity of the $\gamma$-ray emission in the 1-10\,TeV energy band is consistent with the upper limits from the \textit{H.E.S.S.} ToO program on monitoring CC-SNe. The (unabsorbed) peak luminosity is reached after $\approx10\,$days.
    \item The unabsorbed luminosity at \textit{Fermi-LAT} energies peaks earlier, but no CC-SN since the start of the \textit{Fermi} mission was close enough that our results would predict a detection.
\end{itemize}

\begin{multicols}{2}
\footnotesize
\bibliographystyle{aa}
\bibliography{References}
\end{multicols}

\end{document}